\documentclass[aps,pra,twocolumn,superscriptaddress,amsmath,amssymb,floats,floatfix,english]{revtex4}
\usepackage{graphicx}
\usepackage{amsmath}
\usepackage{amssymb}
\usepackage{braket}

\usepackage{polski}
\usepackage{babel}
\usepackage[utf8]{inputenc}
\usepackage[T1]{fontenc}
\usepackage{ifthen}
\usepackage{times}
\usepackage{bm}
\usepackage{pstricks}
\usepackage{times}
\usepackage{soul}
\usepackage{color}
\usepackage{placeins}

\newcommand{\out}[1]{} 							
 
\begin{document}
\title{
Classical limit of entangled states of two angular momenta}
\author{Marek Ku\'s}
\affiliation{Center for Theoretical Physics, Al. Lotnik{\'o}w 32/46, 02-668 Warszawa, Poland
}
\email{marek.kus@cft.edu.pl}

\author{Jan Mostowski}
\affiliation{Institute of Physics of the Polish Academy of Sciences, Al. Lotnik{\'o}w 32/46, 02-668 Warszawa, Poland
}
\email{mosto@ifpan.edu.pl, pietras@ifpan.edu.pl}
\author{Joanna Pietraszewicz}

\affiliation{Institute of Physics of the Polish Academy of Sciences, Al. Lotnik{\'o}w 32/46, 02-668 Warszawa, Poland
}

\date{\today}
\begin{abstract}
We consider  a system of two particles, each with large angular momentum $j$, in the singlet state. 
The probabilities of finding projections of the angular momenta on selected axes are determined.
The generalized Bell inequalities involve these probabilities and we study them using statistical methods. We show that most of Bell’s inequalities cannot be violated, or are violated only marginally, in the limit $j\to \infty$. The precision required to confirm a violation appears to be difficult to achieve. In practice, the  quantum system, in spite of being entangled, becomes indistinguishable from its classical counterpart.
\end{abstract}

\maketitle

\section{Introduction}
\label{sec_introduction}

Large quantum system should exhibit classical features. This is the main idea of the correspondence principle formulated by Bohr in the first years of quantum mechanics. There still is a question how this classical behavior is reached. This applies in particular to systems in entangled states. Quantum system in an entangled state consists of two subsystems, and coherence between these subsystems is the essence  of entanglement.

The classical limit has been studied via different routes. One universal mechanism of reaching the classical limit by a large quantum system is through interactions with environment. The interactions lead to "dephasing" of the quantum system -- during time evolution the density matrix evolves from describing a pure state to a mixed state, an incoherent superposition of special "pointer" states \cite{Zurek_1981, Zurek_2005,Venugopalan_1999}.

Another possibility of discussing the classical limit is to consider quantum mechanical quantities
that have their counterparts in classical physics. Examples of such quantities are expectation values of position, momentum, energy, etc. Moreover, off-diagonal elements of position or momentum operators between states of different energy have their classical counterparts as well.

Yet another aspect of classicality to notice is that very precise measurements have to be performed in order to detect quantum effects in systems close to the classical limit.

Bell \cite{Bell}  found a set of inequalities that are fulfilled by probabilities, obtained within any hidden variable theory with local realism. Violation of Bell's inequalities is a proof that quantum mechanics cannot be replaced by any form of probabilistic theory with local realism. This assertion was confirmed by experiments with two state systems (qubits).

 Original and early versions of Bell's inequalities \cite{Bell,Clauser} involved two observers, each one having a choice of two mutually incompatible experiments. The inequalities have been generalized and can involve many observers, many particles~ \cite{Cabello_2002} and multi-dimensional systems \cite{Collins_2002,Chen13,Popescu,Kaszlikowski,Pal10}. From a geometric point of view, Bell's inequalities describe a bound convex set -- intersection of a finite number of half spaces. Finding all Bell's inequalities gives a necessary and sufficient condition for deciding whether a given state can be viewed as representing a local theory with hidden variables. This is, however, a~ computationally demanding NP~ problem~ \cite{Avis}. A complete list of Bell's inequalities exists only for the simplest cases \cite{Rosset,Clauser,Pironio,Pitovsky,Sliwa1,Gisin19}. 

 More past studies \cite{Garg_1982,Ardehali_1991,Gisin} were interested in the $j\to \infty$ limit.
 They mostly referred to the question of how fast the maximum magnitude of violation vanishes for the optimum angle. The conclusion was that in the nonseparable singlet state  \cite{Garg_1982} large quantum numbers are no guarantee of classical behavior. This view was repeated by \cite{Collins_2002,Chen13}  and was also confirmed using the "numerical linear optimization" approach  \cite{Kaszlikowski, Popescu}. In fact, this approach assumes the non ideal Bell state as a density matrix,  but the measurement itself was treated, from assumption, as an  "ideal one". On the basis of such assumptions, a general belief emerged that probabilities of nonlocal events grow with the dimension of the systems \cite{Collins_2002,Chen13}. This means that the probability of the averaged event might grow beyond the shot noise limit, and the chances of the inequality violation for a given experimental setting increases.

 In this paper we are interested in the classical limit of entangled states, but not from the point of view of robustness to noise of a perfectly chosen Bell inequality. Our effort is to consider the chance to observe a violation when the measured quantity or Bell coefficient settings are imperfectly known.
 We focus on a two particle system, each particle possessing angular momentum $j$, e.g., spinning tops. The classical limit in such a system is reached when the values of both angular momenta are much larger than $\hbar$ -- the quantum unit of angular momentum. We will show that it is very hard to find Bell's inequality and prove the quantum nature of the entangled state when $j \to \infty$. This illustrates the fact that the system cannot be distinguished, or at least it would be very hard to distinguish, from a classical system of two classical correlated angular momenta. 

This approach to the classical limit sheds more light on the structure of entanglement of systems close to the classical limit. Moreover, our view can be quite well suited to experimental conditions, where angles measurements are imperfect and determination of generalized Bell's coefficients for large $j$ is onerous.

\section{
The system:\quad Quantum and classical description
}
\label{system}
We consider a system consisting of two particles (tops), each characterized by angular momentum quantum number~ $j$. The square of the angular momentum is thus $\hbar^2j(j+1)$. We are interested in case of large $j$, hence we consider states close to the classical limit. We assume that angular momenta are expressed in units of $\hbar$. The system is in the state with the total angular momentum equal to zero. This state is given by
\begin{equation}
|\Psi\rangle=\sum_{m=-j}^j (-1)^{j-m}\frac{1}{\sqrt{2j+1}}|m\rangle |-m\rangle.
\end{equation}
This is definitely an entangled state with maximal entanglement.

Two observers, say A and B, measure components of angular momenta along arbitrarily chosen axes in the state\,  $|\Psi\rangle$. Let us name these axes\,  ${\bf a}$ for observer\,  A and \, ${\bf b}$
 for observer\,  B. The result of such measurement is a number $m_1$ in case of observer A and $m_2$ in
case of observer B, with $-j\le m_{1,2} \le j$. Thus there are\,  $2j+1$\,  possible outcomes of measurement for each observer. In case of the state $|\Psi\rangle$ with zero total angular momentum the distribution of the results is flat, the probability of finding the value $m$ of the angular momentum by each observer along each axis is $\frac{1}{2j+1}$.

We will now find the probability amplitude $a({\bf a},{\bf b}, m_1,m_2)$ of detecting a value $m_1$
of the angular momentum of the first particle in the direction ${\bf a}$ and a value of $m_2$ of the angular momentum of the second particle in the direction ${\bf b}$. We make use of the fact that
the state $|\Psi\rangle$ is rotationally invariant, i.e. has the same form in all coordinate systems. We choose, therefore, the system of coordinates where vector ${\bf a}$ is along the $z$ axis and vector
${\bf b}$ lies in the $x-z$ plane. Thus the $y$ axis is perpendicular to the plane spanned by the two vectors. The probability amplitude $a({\bf a},{\bf b},m_1,m_2)$ of detecting values $m_1$ and $m_2$ in the chosen  coordinate system is:
\begin{equation}
\begin{split}
& a({\bf a},{\bf b},m_1,m_2)  = \\
& \frac{1}{\sqrt{2j+1}}\sum_{m=-j}^j \sum_{m'=-j}^j(-1)^{j-m}\langle m_1|m\rangle\langle m_2|m'\rangle\  {\bf d}_{-m,m'}^j(\beta),
\end{split}
\end{equation}
where $\beta$ is the angle between ${\bf a}$ and ${\bf b}$. After short manipulations one can write:
\begin{equation}
a({\bf a},{\bf b},m_1,m_2) = \frac{1}{\sqrt{2j+1}}\, (-1)^{j-m_1}\ {\bf d}_{-m_1,m_2}^j(\beta).
\label{quantum amplitude}
\end{equation}
where\,   ${\bf d}_{-m,m'}^j(\beta)$\,  denotes the Wigner rotation function, \cite{Wignerbook}.

\begin{figure}[tb]
\includegraphics[width=0.45\columnwidth]{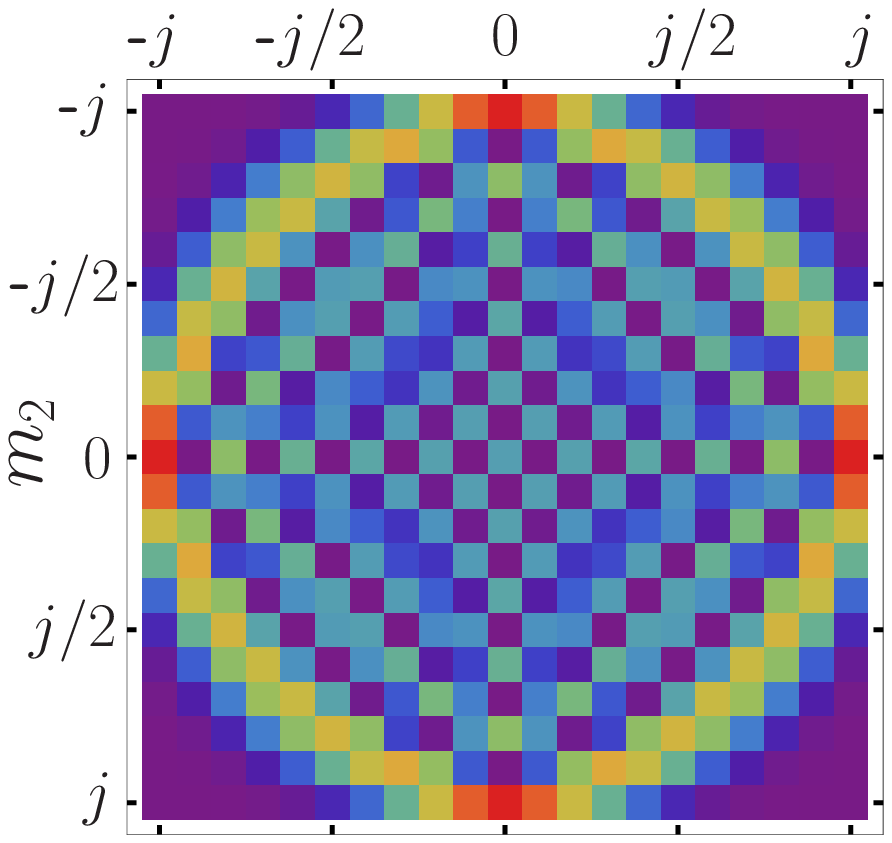}
\includegraphics[width=0.49\columnwidth]{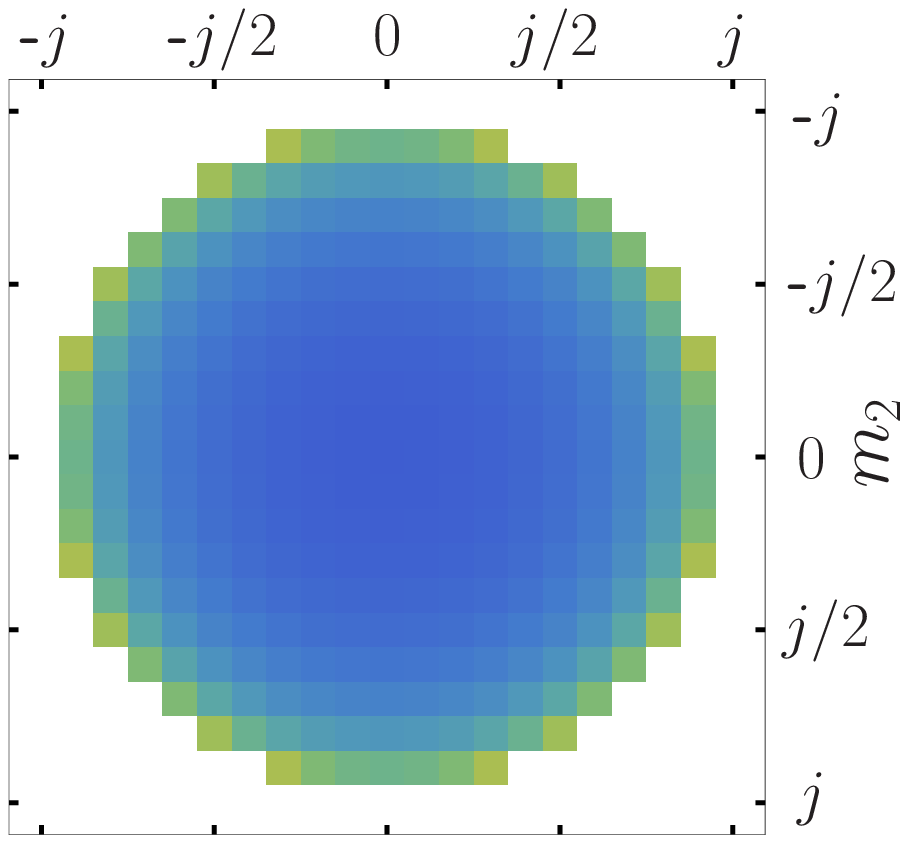}\\
\includegraphics[width=0.45\columnwidth]{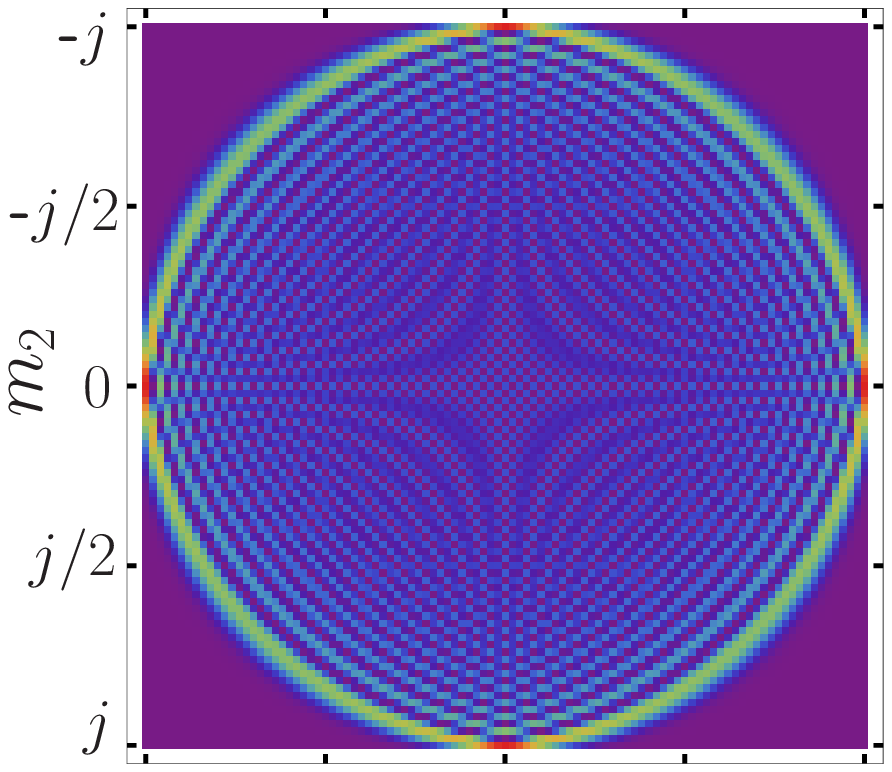}
\includegraphics[width=0.49\columnwidth]{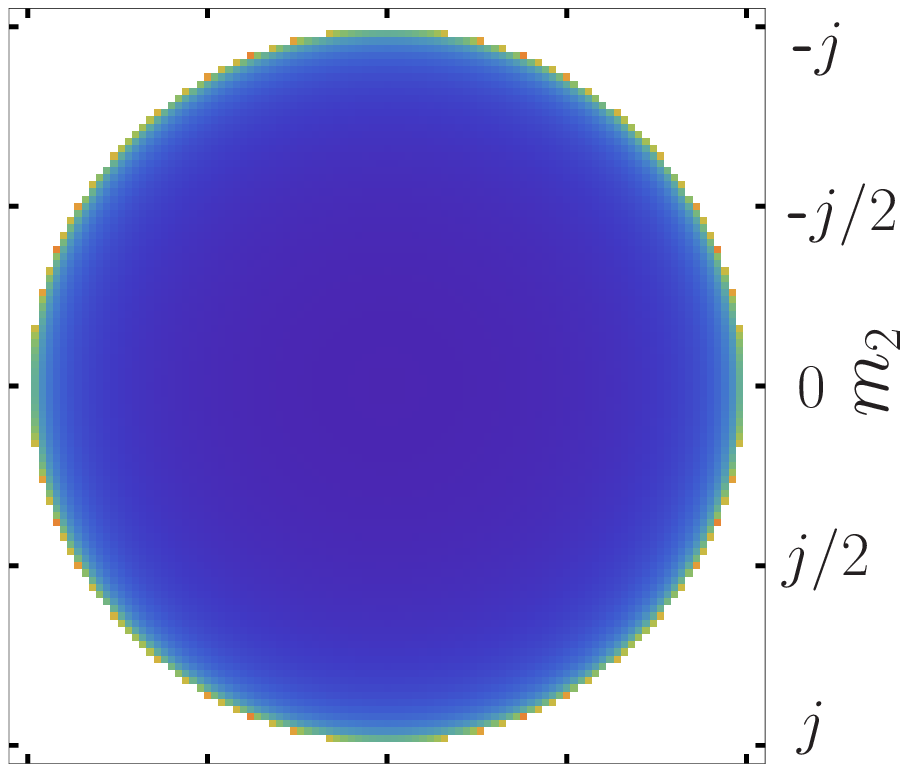}\\
\includegraphics[width=0.45\columnwidth]{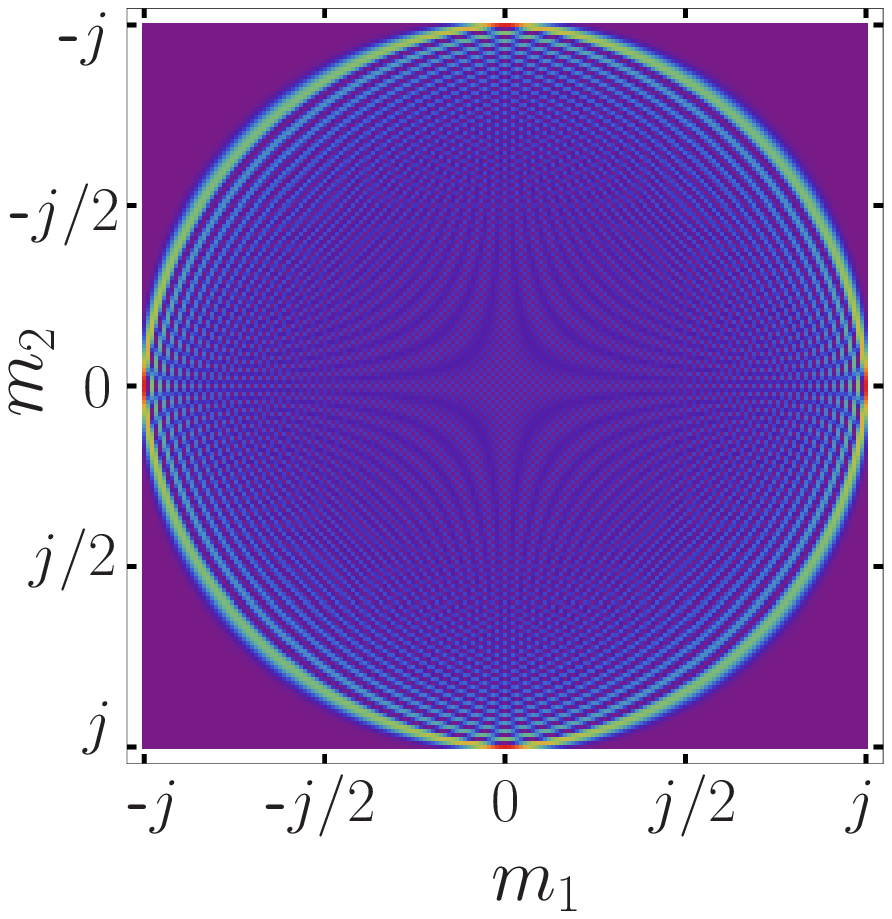}
\includegraphics[width=0.49\columnwidth]{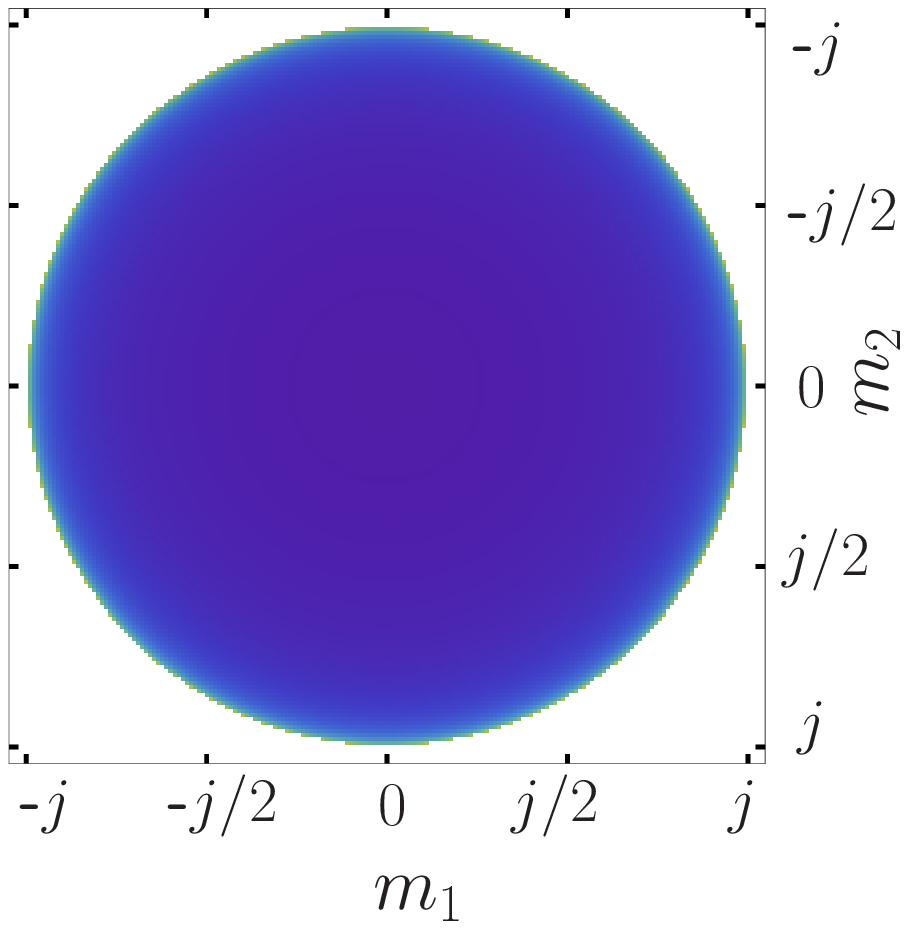}
\caption{
Slices of the quantum (left) and classical (middle and right) probability distribution $|a({\bf a},{\bf b},m_1,m_2)|^2$ when the azimuthal angle between vectors $\bf a$ and $\bf b$ is equal $\beta=\pi/2$. Upper panels: $|j|~=~10$, middle panels $|j|~=~50$, low panels $|j|=90$. Left and middle column plots present the arrayplot function, while right column plots present the continuous density plot.
}
\label{fig:fig1}
\end{figure}

Probability of detecting angular momenta equal to $m_1$ and $m_2$ along appropriate directions is plotted in Fig.\ref{fig:fig1} for several values of $j$. Let us note that this probability depends only on the angle between vectors ${\bf a}$ and ${\bf b}$. Because of rotational invariance of the state $|\Psi\rangle$, the same formula is valid in all coordinate systems.

In addition to the probability amplitude $a({\bf a},{\bf b},m_1,m_2)$ given by (\ref{quantum amplitude}), we will consider the lowest order correlation functions of two angular momentum components along an arbitrary axis $\bf a$ of one angular momentum and the second angular momentum component along a different axis $\bf b$. This correlation function $E({\bf a}\, {\bf J}_1,{\bf b}\, {\bf J}_2)$
is obtained by taking the expectation value of the operator\  ${\bf a}\, {\bf J}_1\cdot{\bf b}\, {\bf J}_2$ in the state $|\Psi\rangle$, where ${\bf J}_1$,${\bf J}_2$ denote angular momentum operators corresponding to the first and second particle. We choose the ${\bf a}$ vector along the $z$ axis and get: 
\begin{equation}
E({\bf a}\, {\bf J}_1,{\bf b}\, {\bf J}_2)=\sum_{m=-j}^j \frac{1}{2j+1} m(-m)b_z.
\end{equation}
Quantity $b_z$ is now the projection of the ${\bf b}$ vector on the $z$ axis, hence the cosine of the angle between vectors ${\bf a}$ and ${\bf b}$. The sum over $m$ is well known. In general, we can write:
\begin{equation}
E({\bf a}\, {\bf J}_1,{\bf b}\, {\bf J}_2)=-\frac{j(j+1)}{3}\, {\bf a} \cdot{\bf b}.
\label{corrquantum}
\end{equation}
Note that (\ref{corrquantum}) is consistent with the correlation function found in the case of spin $\frac{1}{2}$ system (see \cite{Peresbook}).

We will proceed with the study of a purely classical model of two correlated systems having angular momentum $J$ each. The systems are spatially separated, however, their total angular momentum is zero. The system may consist of two spinning tops, with opposite axes of rotation. The direction of the axes is random with uniform probability distribution, therefore the average value of angular momentum of each top is equal to zero. Random uniform distribution of the rotation axis means that its direction ${\bf n}$ is an unit vector, while its polar angle $\Theta$ and azimuthal angle~ $\Phi$ are random numbers with distribution $\frac{1}{4\pi}\sin \Theta$.

We will determine the probability distribution $\rho({\bf a},{\bf b},K,L)$ of finding the value $K$ of the component along the ${\bf a}$ axis of first angular momentum  ${\bf J}_1$,  and the value $L$ of the second angular momentum  ${\bf J}_2$ along the ${\bf b}$ axis. This is the classical counterpart of the quantum expression for $|a({\bf a},{\bf b},m_1,m_2)|^2$\, in (\ref{quantum amplitude}). We will use the fact that the vector ${\bf J}_1$ of the first top has direction along vector ${\bf n}$ and the vector ${\bf J}_2$ of the second top has direction $-{\bf n}$, i.e. ${\bf J}_1={\bf n}J $, ${\bf J}_2=-{\bf n}J$. The classical probability distribution is then given by the formula
\begin{equation}
R({\bf a},{\bf b},K,L)=\frac{1}{4\pi}\int d\Phi\, \sin\Theta\,  d\Theta\, \  \delta(K-J{\bf a}\, {\bf n})\cdot\delta(L+J {\bf b}\, {\bf n}).
\label{classical probability}
\end{equation}
Explicit evaluation of this integral leads to:
\begin{equation}
R({\bf a},{\bf b},K,L)=\frac{1}{2\pi J}(-K^2-L^2-2\, K L \cos\theta +J^2\, \sin^2\theta)^{-\frac{1}{2}},
\end{equation}
where $\theta$\,  is the angle between vectors ${\bf a}$\,  and\,  ${\bf b}$. Let us introduce  scaled angular momenta $k=K/J$ and $l=L/J$. The probability of
 finding the quantities $k$ and $l$
within  the  range of $dk$  and~ $dl$  is simply expressed by $p_c({\bf a},{\bf b},k,l)~=~\rho({\bf a},{\bf b},k,l)\,  dk\, dl$.  The probability distribution therefore reads 
\begin{equation}
\rho({\bf a},{\bf b},k,l)=\frac{1}{2\pi}(-k^2-l^2-2\, k\, l \cos\theta +\sin^2 \theta)^{-\frac{1}{2}}.
\end{equation}

In order to compare quantum probabilities with their classical counterparts, we choose the classical angular momentum\,  $J$\,  equal to the quantum angular momentum $\hbar\sqrt{j(j+1)}$. Also the classical projections $k$ and $l$ are identified with\  $m_1/j$ and\ $m_2/j$. The probability $p_c({\bf a},{\bf b},k,l)$, with\,  $dk=1/j=dl$  is plotted in Fig.~\ref{fig:fig2}.

Now we will discuss the classical correlation $E({\bf a}\, {\bf J}_1,{\bf b}\, {\bf J}_2)$ between a component of one angular momentum along an arbitrary axis $\bf a$  and of the second angular momentum component along a different axis $\bf b$. The correlation is defined as the average value of the product  $({\bf a}\, {\bf J}_1\cdot {\bf b}\, {\bf J}_2$), according to:
\begin{equation}
E({\bf a}\, {\bf J}_1,{\bf b}\, {\bf J}_2)=
-\frac{L^2}{4\pi}\int \, d\Theta\,  \sin\Theta\,  d\Phi\,  ({\bf a}\cdot{\bf n})({\bf b}\cdot{\bf n}).
\end{equation}
Calculation of the integral gives:
\begin{equation}
E({\bf a}\, {\bf J}_1,{\bf b}\, {\bf J}_2)=-\frac{1}{3}J^2\ {\bf a}\cdot{\bf b}.
\label{corrclass}
\end{equation}
Correlation (\ref{corrclass}) found in this purely classical limit is consistent with the quantum correlation (\ref{corrquantum}), the only difference is that there is $J^2$ in the classical case as opposed to\  $\hbar^2 j(j+1)$ in the quantum case.

\section{Semiclassical approximation}
\label{semapprox}
Semiclassical approximation to the quantum description provides a link between the quantum and classical approaches. We will find now the semiclassical approximation to the probability amplitude $a({\bf a},{\bf b},m_1,m_2)$ and discuss its relation to classical probability. We will use the WKB method treating $\frac{1}{j}$ as a small parameter.

We will start with the full Wigner rotation function ${\bf D}_{m_1,-m_2}^j(\alpha,\beta,\gamma)$ \cite{Wignerbook}. It satisfies the differential equation
\begin{equation}
\begin{split}
\Big[-\frac{1}{\sin^2\beta}\left(\frac{\partial^2}{\partial\alpha^2}+\frac{\partial^2}{\partial\gamma^2}  2\cos\beta\frac{\partial^2}{\partial\alpha\partial\gamma}\right)-
\frac{\partial^2}{\partial\beta^2}- \\
\cot\beta\frac{\partial}{\partial\beta} j(j+1)\Big]\, D_{m',m}^j(\alpha,\beta,\gamma)=0.
\end{split}
\end{equation}

The standard semiclassical (WKB) method allows one to find the asymptotic behavior of the ${\bf D}_{m,m^\prime}^j(0,\beta,0)={\bf d}_{m,m^\prime}^j(\beta)$, where ${\bf d}_{m,m^\prime}^j(\beta)$\  is \,  the\,   matrix  element of\,  the\,  rotation\,  around the~ $y$-axis~ \cite{Braun96}:
\begin{equation}\label{dmm0}
{\bf d}_{m,m^\prime}^j(\beta)=(-1)^j\sqrt {\frac{{2\bar J}}{\pi }\left| {\frac{{\partial ^2 S_0 }}{{\partial m\, \partial m'}}} \right|}\, \cos{\left(\bar J\, S_0(\beta)  - \frac{\pi }{4} \right)},
\end{equation}
where ${\bar J}=j+\frac{1}{2}$ and $S_0$ is the generating function of the classical rotation:
\begin{eqnarray}
\label{s0}
S_0 \left( {m',m} \right) &=& \frac{m}{{\bar J}}\arccos\left(\frac{{m\cos \beta  - m'}}{{\sin \beta \sqrt {\bar J^2  - m^2 } }}\right) \nonumber \\
&-& \frac{{m'}}{{\bar J}} \arccos\left(\frac{{m - m'\cos \beta }}{{\sin \beta \sqrt{\bar J^2  - m'^2 } }}\right) \nonumber \\
&+& \arccos\left( \frac{{m\, m' - \bar J^2 \cos \beta }}{{\sqrt {\left( {\bar J^2  - m'^2 } \right)\left( {\bar J^2  - m^2 } \right)} }}\right) \quad
\end{eqnarray}
Hence,
\begin{eqnarray}\label{key}
&&{\bf d}_{m,m^\prime}^j(\beta)=(-1)^j\cos \left( {\bar JS_0(\beta)  - \frac{\pi }{4}}\right) \times
\nonumber \\
&&\times
\left( {\frac{2}{{\pi \sqrt {\sin ^2 \beta  - \frac{1}{{\bar J^2 }}\left( {m^2  + m'^2  - 2\, m\, m'\, \cos \beta } \right)} }}} \right)^{\frac{1}{2}}. \nonumber \\
\label{phrase}
\end{eqnarray}

These approximate formulas are not valid in the vicinity of the classical "turning points", i.e points where the denominator in  (\ref{phrase}) is close to zero.

\begin{figure}[t]
\includegraphics[width=0.45\columnwidth]{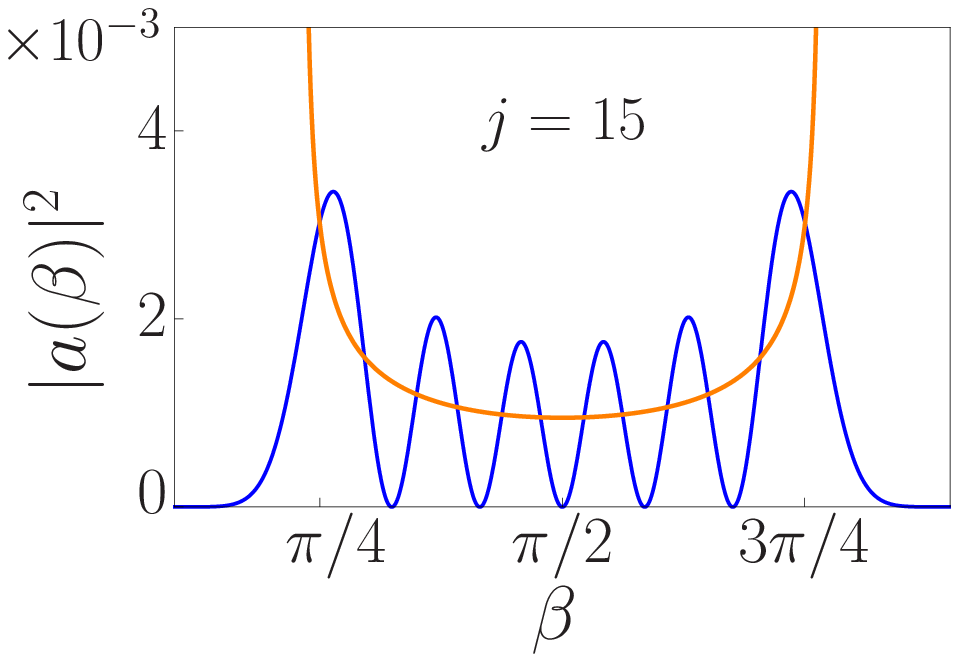}
\includegraphics[width=0.45\columnwidth]{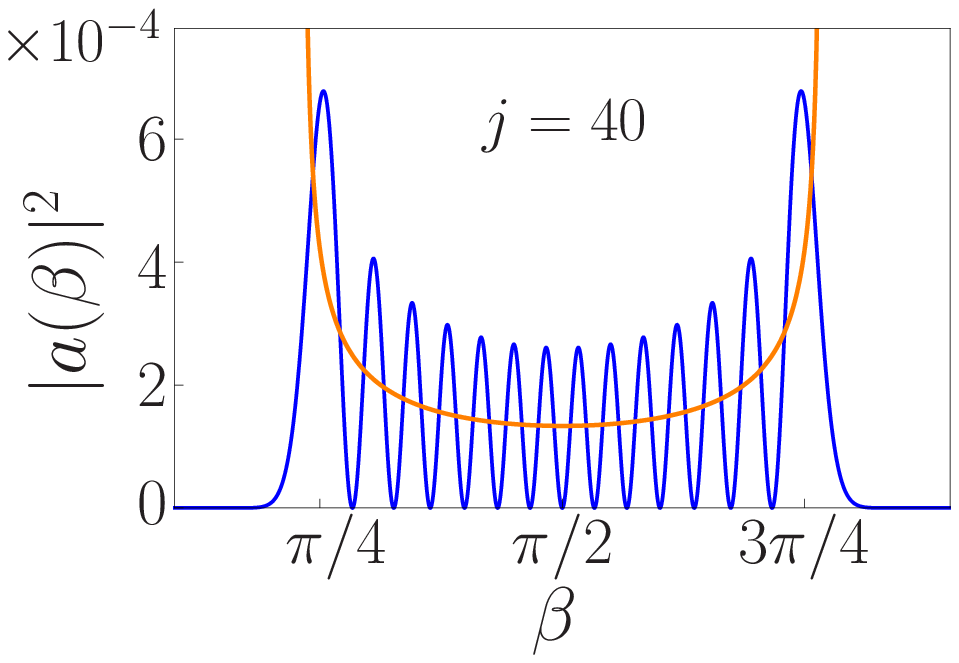}\\
\includegraphics[width=0.45\columnwidth]{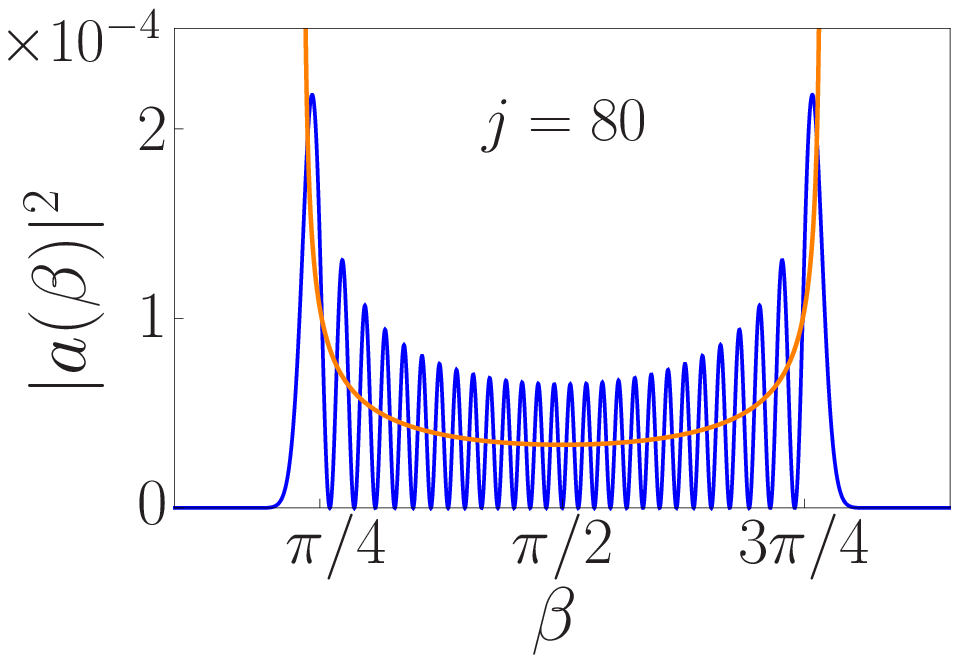}
\includegraphics[width=0.45\columnwidth]{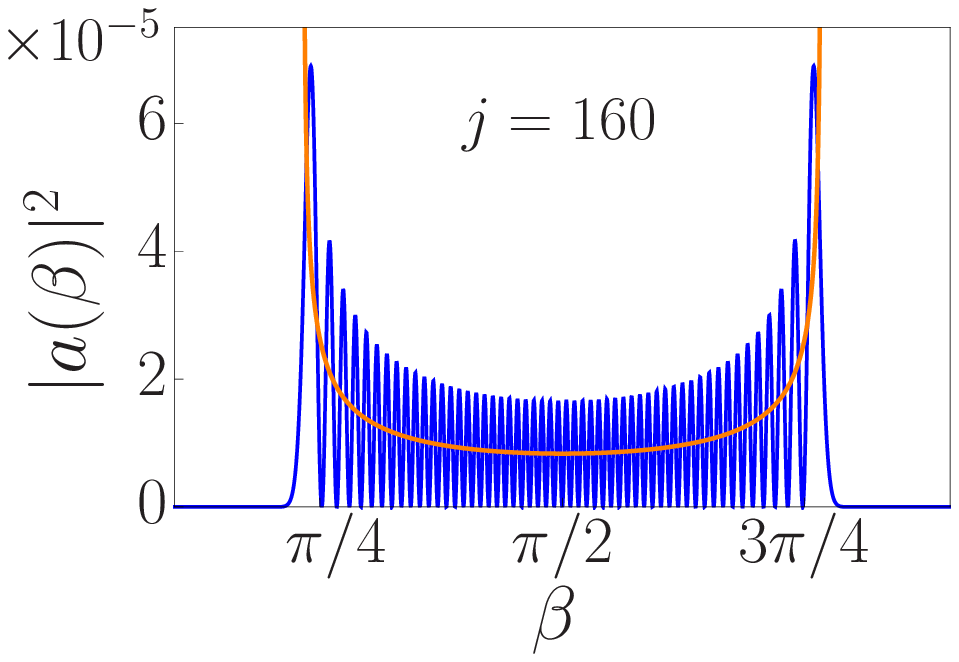}
\caption{
 Probability distributions as a function of the projected angle $\beta$ for various momenta\ $j$ (values on the plots). The classical ratios\, $k/j= 0.1$ and $l/j= -0.66$ \, are fixed,  while the quantum  ones, namely $m_1$ and $m_2$,  are the nearest values allowed by integer $n = 2$. The blue line corresponds to quantum probability $|a({\bf a},{\bf b},m_1,m_2)|^2=|a(\beta)|^2$ and the orange line corresponds to classical probability distribution $\rho({\bf a},{\bf b},k,l)$. 
}
\label{fig:fig2}
\end{figure}

Comparison of exact values of the ${\bf d}_{m',m}^j(\beta)$ functions and their semiclassical approximations is illustrated in~Fig.\ref{fig:fig2}. Observe, that the probability distribution $\rho({\bf a},{\bf b},k,l)$ found in the previous section is equal to the square of the envelope
of the ${\bf d}_{m,m'}^j(\beta)$ function in the semiclassical approximation. The difference between the studied probability of the quantum system and the corresponding probability distribution for the classical system lies in small high frequency oscillations seen in Fig.~\ref{fig:fig2}. 

\section{Statistical approach to generalized Bell's inequalities}

In this section we will concentrate on entanglement of the system and possible experimental tests that can verify its existence. 
Entanglement is a nonclassical feature of quantum systems. The question arises as to whether entanglement can be detected in a system that is close to the classical limit. 
States can be entangled, of course, regardless of the level of excitation. The example of the state studied here proves this beyond any doubt. Regardless of the value of $j$ the state $|\Psi\rangle$ is  maximally entangled  (regardless of exact definition maximal entanglement). The real question is can one prove in experiments that entanglement in this state really exists.

Bell's inequalities and their generalizations provide a powerful tool to prove existence of entanglement. The idea of Bell's inequalities can be formulated as follows. One considers a set of
probabilities:\, $p({\bf a}, m)$ and\, $p({\bf a},{\bf b}, m_1,m_2)$ where the measurement axes are chosen in various directions defined by vectors  ${\bf a}$ and ${\bf b}$. Two sets of vectors ${\bf a}_r$ and ${\bf b}_s$, where $r=1,2$ and $s=1,2$ should be considered. Notice that probabilities $p({\bf a}_r,{\bf b}_s, m_1,m_2)$ depend on vectors ${\bf a}_r$ and ${\bf b}_s$ by the angle between them.
If hidden variables exist and local realism is valid then probabilities\, $p({\bf a}_r, m)$ and\,  $p({\bf a}_r,{\bf b}_s, m_1,m_2)$ are within the convex hull spanned by vertex  defined by these functions. This means that the probabilities satisfy a set of inequalities known as (generalized) Bell's inequalities.

Numerous experiments showed that these inequalities are violated by quantum probabilities in case of entangled states of two $j=1/2$ states, known as qubits \cite{Peresbook}, and references therein. This provides a strong argument that no hidden variables or local realism exist in quantum physics. 

Let us now discuss in more detail the generalized Bell's inequalities. These are inequalities that should be satisfied by linear combinations of joined probabilities $p({\bf a}_r,{\bf b}_s, m_1, m_2)$ under the assumption of local realism and existence of hidden variables. Analogs of Bell's inequalities in case of large angular momentum are also known \cite{Sliwa1, Pitovsky}; see also \cite{Garg_1982, Mermin_1983,Horodecki_1996,Mermin_1996,Mermin_1990,Gisin,Popescu}. 

In case of arbitrary $j$ there are $S=4\times (2j+1)^2$ joined probabilities $p({\bf a}_r,{\bf b}_s, m_1, m_2)$. According to \cite{Peres}, \ all Bell's inequalities are of the form
\begin{equation}
M_1\le \sum_{r,s,\mu,\nu} p({\bf a}_r,{\bf b}_s,m_\mu,m_\nu) \, c(r,s,m_\mu,m_\nu)\le M_2,
\label{ineq}
\end{equation}
where $M_1$ and $M_2$ are fixed numbers and  $c({r,s,m_\mu,m_\nu})$\, are coefficients. The coefficients are equal to 0 or to natural numbers with plus or minus sign, \cite{Peres}. Not all of them are independent, however, the number of independent inequalities (\ref{ineq}) grows rapidly with $j$. Since we are interested in the case of large $j$, the number of Bell's inequalities is huge. It is not realistic to consider them all, so we will turn to the statistical approach.

Inspiration to use statistical methods stems from a somewhat similar approach to spectra of complex systems, where the Hamilton operator is replaced by a matrix with elements being random numbers with a statistical distribution. The eigenvalues are random numbers and in fact are studied using statistical methods.

\begin{figure}[b]
\includegraphics[width=0.85\columnwidth]{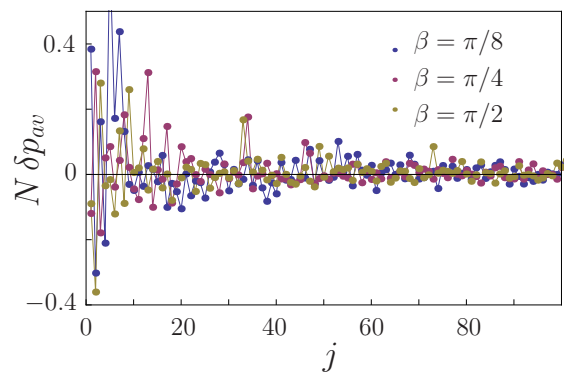}
\caption{The sum of  all quantum corrections versus momentum $j$ for a given set of angle $\beta$. Color lines-points correspond to different values of angle $\beta$.}
\label{fig:averdp}
\end{figure}

To deal with Bell's  inequalities, we will treat the probabilities of measuring given values of angular momenta by the two observers as random numbers. Quantum probabilities $p({\bf a}_r,{\bf b}_s,m_1, m_2)$ have their classical analog $p_c({\bf a}_r,{\bf b}_s,k,l) $, as we saw in Sec.\ref{semapprox}. 
It is helpful to consider quantum corrections, i.e. differences between the quantum and classical probabilities. The quantum corrections are
\begin{equation}
\delta p({\bf a}_r,{\bf b}_s, m_1,m_2)=p({\bf a}_r,{\bf b}_s,m_1,m_2)-p_c({\bf a}_r,{\bf b}_s,k,l)dk dl, \end{equation}
where $k=m_1/j$, $l=m_2/j$ and increments are chosen to fit the increments of quantum numbers $m_1$ and $m_2$, therefore $dk = 1/j = dl$.

Classical probabilities satisfy Bell's inequalities and violation of these inequalities by quantum systems can be due to the quantum corrections only.

Now we will study how the corrections scale with the value of angular momentum $j$. First we will show that the corrections \, $\delta p({\bf a}_r,{\bf b}_s,m_1,m_2)$ tend to zero for large\,  $j$ regardless of the choice of vectors ${\bf a}_r$~ and~ ${\bf b}_s$. Since there are many corrections mentioned above, we will study average values of them, i.e.\, $\delta p_{av}~=~\frac{1}{(2j+1)^2}\sum_{m_1,m_2}\delta p({\bf a}_r,{\bf b}_s,m_1,m_2)$.
\begin{figure}[t]
\includegraphics[width=0.85\columnwidth]{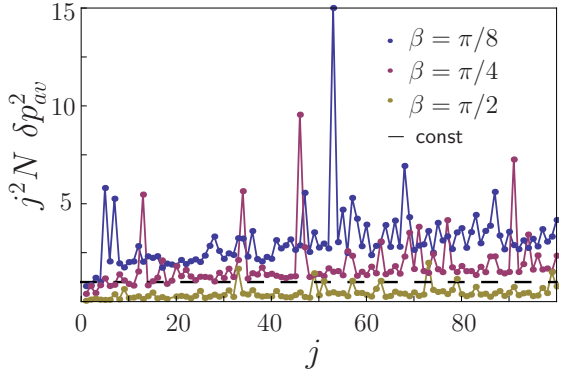}
\caption{
 The scaling of  average squares of $\delta p({\bf a}_r, {\bf b}_s, m_1,m_2)$ versus momentum $j$. The scaling factor is proportional to~ $j^2N$, where $N~=~(2j+1)^2$. 
}
\label{fig:averdp2}
\end{figure}
The behavior of this quantity is  illustrated in Fig.\ref{fig:averdp} as a function of $j$. Notice that $\delta p_{av}$ tends to zero when\,  $j\rightarrow\infty$. The next figure, Fig.\ref{fig:averdp2},  shows the average squares of\, $\delta p_{av}$, i.e. 
$\delta p^2_{av}~=~\frac{1}{(2j+1)^2}\sum_{m_1,m_2}(\delta p({\bf a}_r, {\bf b}_s,m_1,m_2))^2$ as functions of $j$.
It is clear that the larger the $j$ the smaller the average difference between the classical and quantum probability. Notice from Fig.\ref{fig:averdp2}\,  that\,  $\delta p^2_{av}$ goes to zero as \, $j^{-4}$.

\begin{figure}[t]
\includegraphics[width=0.95\columnwidth]{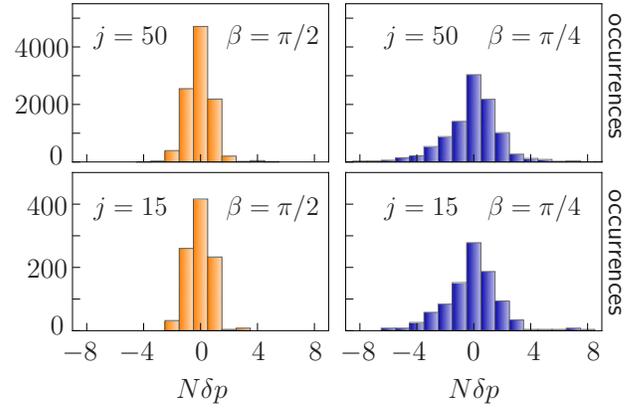}
\caption{
Histograms of \ $N\, \delta p({\bf a}_r,{\bf b}_s,m_1,m_2)$
values, which are given by  projections of momentum $m_{1,2}$ and angle $\beta$ between vectors ${\bf a}_r$ and~ ${\bf b}_s$.  Values $m_{1,2}$ vary from $-j$ to $j$ by one. Values $j$ and $\beta$ are placed on the plot.
}
\label{fig:histo}
\end{figure}

Figure \ref{fig:histo} presents histograms of differences between classical and quantum probabilities for various $j$ and various angles $\beta$ between vectors ${\bf a}_r$ and ${\bf b}_s$. We see that the differences are localized around zero and the distribution is close to a Gaussian. Thus we may say that the differences between classical and quantum probabilities are random numbers with a narrow distribution.

\begin{figure}[b]
\includegraphics[width=0.85\columnwidth]{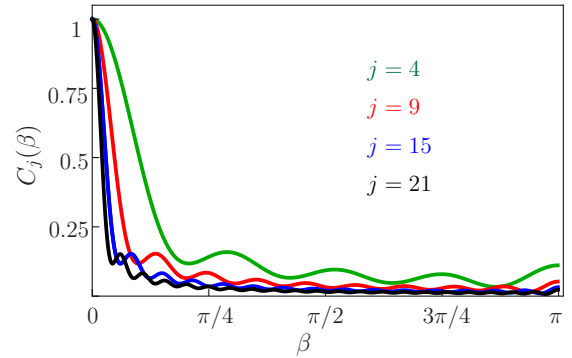}
\caption{
Correlation $C_j(\beta)$ as a function of angle between vectors ${\bf a}_r$ and ${\bf b}_s$ for various momentum $j=4$ (green), $j=9$ (red), $j=15$ (blue), $j=21$~ (black).
}
\label{fig:correl}
\end{figure}

Finally we will show that probabilities $p({\bf a}_r,{\bf b}_s,m_1,m_2)$ for various angle between ${\bf a}_r$ and ${\bf b}_s$ are not correlated. This is visible in Fig.\ref{fig:correl}, where the correlation function: 
\[
C_j(\beta)=N\sum_{m_1,m_2=-j}^j\left[ p({\bf a}_r,{\bf a}_s,m_1,m_2)\, \cdot \, p({\bf a}_r,{\bf b}_s,m_1,m_2) \right]
\] 
decreases fast with $j$ in the vicinity of small angles.

The discussion of Bell's inequalities (\ref{ineq}) will be continued using differences $\delta p^2_{av}$. The generalized inequalities (\ref{ineq}) involve linear combinations of probabilities. We can subtract from them the classical values, that obviously fulfill the inequalities, and get a set of inequalities for $\delta p({\bf a}_r,{\bf b}_s,m_1,m_2)$.  Every such inequality contains $(2j+1)^2$ terms at most, so the number of terms scales as $j^2$.  Note that each term consists of a product of a coefficient\,  $c(r,s,m_1,m_2)$\,  and\,  $\delta p({\bf a}_r,{\bf b}_s,m_1,m_2)$. The number of coefficients grows with~ $j^2$, while $\delta p({\bf a}_r,{\bf b}_s,m_1,m_2)$  tend to zero as $j^{-2}$.
Therefore the linear combination can either tend to zero or to a constant, independent of $j$. Since $\delta p({\bf a}_r,{\bf b}_s, m_1,m_2)$ can be positive or negative the linear combination of them in any of Bell's inequalities contains terms with alternating signs and thus the total sum tends to zero with $j\rightarrow\infty$. The only exception is the case when the signs of $\delta p({\bf a}_r,{\bf b}_s,m_1,m_2)$ are correlated with the signs of $c({r,s,m_1,m_2})$, i.e. when all or at least most products\ $c(r,s,m_1,m_2)\times \delta p({\bf a}_r,{\bf b}_s,m_1,m_2)$\  are of the same sign.

We should stress also that probabilities $p({\bf a}_r,{\bf b}_s,m_1,m_2)$ as well as  $\delta p({\bf a}_r,{\bf b}_s,m_1,m_2)$ depend on the choice of measurement performed by the two observers. A measurement is defined by the directions ${\bf a}_r$ and ${\bf b}_s$ along which the angular momenta are measured. From Fig.\ref{fig:fig2} we can clearly see that the probabilities $p({\bf a}_r,{\bf b}_s,m_1,m_2)$ are quite sensitive to the angle between the two vectors. Even a small change of the angle leads to a large change of $p({\bf a}_r,{\bf b}_s,m_1,m_2)$. Suppose one of Bell's inequalities is violated by a set of probabilities for given vectors ${\bf a}_r$ and ${\bf b}_s$. A small change of these vectors, or rather the angle between them, by about $\pi/j$ leads to a substantial change of the probabilities. In fact, instead of minima of $p$ (as functions of angle) we may have maxima after a small change of angles and the inequality is no longer violated. Thus verification if a Bell inequality is violated or not, depends on the choice of angles, and therefore on the accuracy of angle determination.

In order to strongly violate a Bell inequality the coefficients $c(r,s,m_1,m_2)$ have to be correlated with the probabilities $p({\bf a}_r,{\bf b}_s,m_1,m_2)$. This requires that the $c(r,s,m_1,m_2)$ coefficients corresponding to large probabilities $p({\bf a}_r,{\bf  b}_s,m_1,m_2)$, are large (in absolute value) and negative to dominate over the remaining positive terms. These "specially designed" conditions do not prove that Bell's inequalities cannot be violated by probabilities obtained form the state $|\Psi \rangle$. The scaling laws found above say that on average the violation of Bell's inequalities is negligible. Therefore, possible determination of the violation is not probable. A small number of violated  inequalities indicate that the degree of practical violation does not grow with $j$. In this way the quantum state, $|\Psi \rangle$, being a quantum maximally entangled state, reproduces classical angular momentum. 

Observe that the classical limit proved here is not reached in a uniform way. It is not said that each of Bell's inequalities is fulfilled  in the limit of large $j$.
 
A completely different approach to Bell's inequalities is to consider correlation functions rather than the probabilities, see \cite{Peresbook}. The correlation function $E$ can be written as a sum $E=E_{classical}+E_{quantum}$, where $E_{classical}=-\frac{1}{3} {\bf a}\cdot{\bf b}$, and $ E_{quantum}=-\frac{1}{3j}{\bf a}\cdot{\bf b} $. In the classical limit $j\rightarrow\infty$ 
correlation functions $E$ tend to their classical counterparts and no violation of Bell's inequalities is possible.

\section{Conclusions}

There are many approaches to the study of the classical limit in case of large quantum system. One kind of approach is based on measurements and finite resolution of any measuring device. If the resolution of angular momentum measurement is below the quantum unit  $\hbar$ then one cannot measure the probability $p({\bf a}_r, m)$ nor the joined probability  $p({\bf a}_r,{\bf b}_s, m_1,m_2)$. Each such measurement only gives an average over many $m_1$ and $m_2$. It is clear from the plots that such averaging leads directly to classical probabilities. This is a simple observation.

Another approach of looking at the classical limit, which was formulated by Zurek \cite{Zurek_1981}, stresses the importance of the environment and damping of the quantum coherence by mutual interaction between the system and the environment. Our approach to the classical limit is, however, very different from this one. There is no need for coherence damping in the present approach. The probabilities of measurement and also correlation functions of the quantum theory are sufficiently close to their classical counterparts to make the quantum case look very much the same as the classical system.

Obviously, the quantum states differ from the classical distribution function. The difference lies in entanglement that has no classical counterpart. In case of the entangled state the relation between classical and quantum description is  more subtle than in previously mentioned cases. Existence of entanglement can be verified on the basis of Bell's inequalities and their generalizations. While a lot is known about Bell's inequalities for small systems, two-state systems in particular, the case of the large entangled system is less known. We formulated a statistical approach to the problem based on the fact the number of Bell's inequalities grows rapidly with the size of the system and that the probability of measuring a single state is very difficult to determine. Using the example of two large but entangled angular momenta we showed that joined probabilities\,  $p({\bf a}_r,{\bf b}_s, m_1,m_2)$ \, behave similar to random numbers in case of large $j$.

Our statistical approach showed that it is very difficult to find an inequality that is violated by the
entangled state of two large angular momenta. This does not prove that such an inequality does not exist. Our result shows only that there are very few of such violated inequalities and therefore it is very difficult to find them. We should stress that an inequality exists that is violated even in the case of very large systems \cite{Gisin}.

In addition to Bell's inequalities we studied quantum correlation functions and showed that they tend to the classical limit of two correlated classical tops like $1/j$. Because of this rapid approach to the classical limit the correlation functions cannot be used efficiently to discriminate between classical and quantum states.

\section{Acknowledgement}
JP and MK acknowledge support from the National Science Centre (Poland) grants No. 2012/07/E/ST2/01389 and No. 2017/27/B/ST2/0295, respectively.
\bibliography{dist}
\end{document}